\documentclass[aip,rsi,reprint,graphicx]{revtex4-1} % for checking your page length
\usepackage{graphicx}
\usepackage{amsmath}
\usepackage{amssymb}
\usepackage{subfigure}
\usepackage{bm}
\graphicspath{{./figures/}}
\DeclareGraphicsExtensions{.pdf,.png,.jpg,.eps}
\draft % marks overfull lines with a black rule on the right
\begin{document}

% Use the \preprint command to place your local institutional report number
% on the title page in preprint mode.
% Multiple \preprint commands are allowed.
%\preprint{}

\title{Angular instability in high optical power suspended cavities} %Title of paper

% repeat the \author .. \affiliation  etc. as needed
% \email, \thanks, \homepage, \altaffiliation all apply to the current author.
% Explanatory text should go in the []'s,
% actual e-mail address or url should go in the {}'s for \email and \homepage.
% Please use the appropriate macro for the type of information

% \affiliation command applies to all authors since the last \affiliation command.
% The \affiliation command should follow the other information.

\author{Jian Liu}
\email[]{jianliugw@gmail.com}
%\homepage[]{Your web page}
%\thanks{}
%\altaffiliation{}
\affiliation{OzGrav-UWA, Department of Physics, University of Western Australia, Crawley, WA 6009, Australia}

\author{Vladimir Bossilkov }
%\email[]{v.bossilkov@gmail.com}
%\homepage[]{Your web page}
%\thanks{}
%\altaffiliation{}
\affiliation{OzGrav-UWA, Department of Physics, University of Western Australia, Crawley, WA 6009, Australia}

\author{C. Blair}
\affiliation{LIGO Livingston Observatory, Livingston, LA 70754, USA}

\author{C. Zhao}
\affiliation{OzGrav-UWA, Department of Physics, University of Western Australia, Crawley, WA 6009, Australia}

\author{L. Ju}
\affiliation{OzGrav-UWA, Department of Physics, University of Western Australia, Crawley, WA 6009, Australia}

\author{D.G.Blair}
\affiliation{OzGrav-UWA, Department of Physics, University of Western Australia, Crawley, WA 6009, Australia}

% Collaboration name, if desired (requires use of superscriptaddress option in \documentclass).
% \noaffiliation is required (may also be used with the \author command).
%\collaboration{}
%\noaffiliation

\date{\today}

\begin{abstract}

Advanced gravitational wave detectors use suspended test masses to form optical resonant cavities for enhancing the detector sensitivity. These cavities store hundreds of kilowatts of coherent light and even higher optical power for future detectors. With such high optical power, the radiation pressure effect inside the cavity creates sufficiently strong coupling between test masses whose dynamics are significantly altered. The dynamics of two independent nearly free masses become a coupled mechanical resonator system. The transfer function of the local control system used for controlling the test masses is modified by the radiation pressure effect. The changes in the transfer function of the local control systems can result in a new type of angular instability which occurs at only 1.3 \% of the Sidles-Sigg instability threshold power. We report experimental results on a 74~m suspended cavity with a few kilowatts of circulating power, for which the power to mass ratio is comparable to the current Advanced LIGO. The radiation pressure effect on the test masses behaves like an additional optical feedback with respect to the local angular control, potentially making the mirror control system unstable. When the local angular control system is optimised for maximum stability margin, the instability threshold power increases from 4~kW to 29~kW. The system behaviour is consistent with our simulation and the power dependent evolution of both the cavity soft and hard mode is observed. We show that this phenomenon is likely to significantly affect proposed gravitational wave detectors that require very high optical power. 

%Optical springs are proposed to be used in laser interferometer gravitational wave detectors for reaching sensitivity below standard quantum limit. In this %paper we demonstrate that optical spring can result in instability of low frequency mode for a high finesse suspended cavity. This effective sets a %threshold power for the stable operation of the cavity. Experiments in a 80m suspend cavity shows that instability can occur under both blue and red %detuned condition and the intra-cavity threshold power is highly depended on the beam location on the mirrors. As for advLIGO, this would not be a problem %unless the cavity locking precision is larger than 1/8000 of its cavity linewidth.
\end{abstract}

% insert suggested PACS numbers in braces on next line

\maketitle %\maketitle must follow title, authors, abstract and \pacs
% Body of paper goes here. Use proper sectioning commands.
% References should be done using the \cite and \label commands
\section{Introduction}
The direct detection of gravitational waves from colliding black holes and neutron stars opened a new window for studying the Universe. The first direct observation of gravitational waves from two merging black holes, named GW150914, was made by two Advanced LIGO (aLIGO) detectors on $\rm 14^{th}$ September  2015~\cite{abbott2016}. This was followed by four more binary black hole merger detections~\cite{gw151226,gw170104,gw170608,gw170814} and the first binary neutron star in-spiral detection~\cite{gw170817}. As the detector sensitivity increases from instrumentation improvements, coming years will bring about a surge in detection of black hole and neutron star mergers, as well as other sources of gravitational waves, including supernovae. The initial detector array comprising two aLIGO~\cite{aasi2015} detectors and Advanced Virgo~\cite{acernese2015} is the start of a global array. The next addition will be the underground, cryogenic gravitational wave detector KAGRA, which is now nearing completion~\cite{somiya2012}. These advanced kilometer scale interferometers all use long Fabry-P\' erot (FP) optical cavities, designed to contain high optical power up to 800 kW~\cite{aasi2015}. 

Current aLIGO has tripled the peak sensitivity and extended the detection bandwidth compared with the initial LIGO. The critical aspect to achieve high sensitivity is the high circulating power. The circulating optical power in the aLIGO FP arm cavities reached 100 kW~\cite{abbott2016} in the second observation run and it is expected to increase to 200 kW in the third observing run due to begin in early 2019. Higher power reduces the quantum shot noise, but increases the radiation pressure in the detector. It also enhances the optomechanical coupling between the intra-cavity light field and the cavity suspended mirrors, leading to various potential instabilities~\cite{zhao2005,zhao2015,evans2015observation} which can threaten the stable operation of the detector.
 
%Three-mode parametric instability\cite{braginsky2001} (PI)has been observed during commissioning of aLIGO and currently sets a limit of about 200 kW \cite{blair2015} on the operating intra-cavity power. In this phenomenon the cavity fundamental optical mode is scattered by the mirror internal acoustic mode into high order optical modes. The optical modes create a radiation pressure force that in turn drives or damps the acoustic mode. Under certain conditions, the amplitude of the acoustic mode can grow exponentially, leading to parametric instability and failure of the interferometer. PI has been studied intensively. It was first observed at the High Optical Power Test Facility (HOPTF) at Gingin, Western Australia \cite{zhao2005,ju2006,zhao2008,zhao2009,zhao2015} and soon after in Advanced LIGO Livingston Observatory in 2014\cite{evans2015}. Different PI control methods have been proposed and several are being tested currently\cite{gras2009,fan2010,miller2011,degallaix2007}.
 
The test masses in advanced gravitational wave detectors must be controled by applying weak forces, to maintain the cavities on resonance.  Control is normally achieved by a combination of `local control' that uses local position sensing and actuation, plus `global control' in which feedback signals are derived from various detector output ports. As a result of increased optical power, radiation pressure forces create strong coupling between the test masses, such that each test mass can no longer be considered in isolation. 

A suspended cavity exhibits different dynamics in the presence of high radiation pressure and is susceptible to Sidles-Sigg (SS) instability~\cite{sidles2006}. It occurs when the optical spring induced negative torsional rigidity exceeds the positive torsional rigidity of the mirror suspension system. Fan $et.al$ observed angular optical rigidity in a suspended cavity in 2009~\cite{fan2009}. SS instability was observed in initial LIGO in 2010~\cite{hirose2010angular}. In 2016, Yutaro $et.al$ reported observation of optical anti-spring effect in a 23~mg mirror FP cavity~\cite{enomoto2016observation}. In these experiments it was shown that SS instability would occur if the radiation pressure force was large enough. Near-planar cavities are more vulnerable to SS instability than near-concentric cavities. Employing a negative cavity $g$-factor allows for maximised intra-cavity power, by minimising susceptibility to SS instability. 
 
In this paper we report a new type of instability induced by the interaction between the local angular control system and radiation pressure, which has been observed on the high power 74~m suspended optical cavity at Gingin, Western Australia. When the cavity was locked using the PDH method~\cite{black2001}, the two mirrors are coupled by radiation pressure force, resulting two cavity modes named soft and hard mode. Radiation pressure force acts as an additional optical feedback, modifying the local mirror control transfer function. The interaction strength was observed to be proportional to the intra-cavity power. When the intra-cavity power reached a certain threshold, the control loops became unstable. This was due to a reduction of phase margin caused by the cavity hard mode shifting to higher frequency. Thus a control loop designed for low power operation was unable to cope with the new degrees of freedom that arose from the optomechanical coupling between the mirrors. 

The conventional SS instability occurs when the cavity soft mode frequency reaches zero, but the instability reported here can happen at a much lower power level if the initial control system design does not consider the interaction between the hard mode and the control loops. It becomes essential that the control loops should be either designed to prevent both cavity soft mode and hard mode instabilities in the first place or to be adaptive to intra-cavity power.

\section{Sidles-Sigg instability model}\label{ssmodelsection}
\begin{figure}[h!]
  \centering
  \includegraphics[width=8cm]{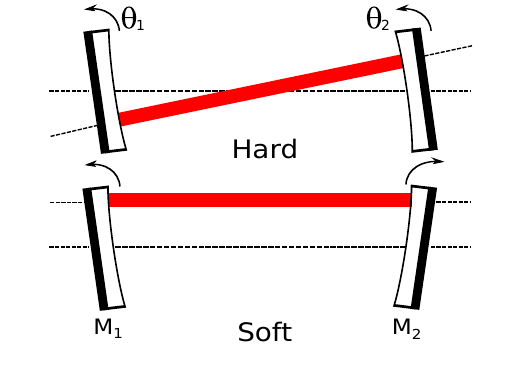}\\
  \caption{Illustration of cavity hard and soft modes in a simple FP cavity composed by two identical suspended mirrors. $\rm{C_{1,2}}$, $\rm{O_{1,2}}$ and $\theta_{1,2}$ are the centre of curvature,  geometric center the rotation angle of mirror $\rm{M_{1}}$ and $\rm{M_{2}}$, respectively. The cavity soft mode in panel (b) is potentially unstable and additional control is needed to maintain stable of the cavity at high level of intra-cavity power.}
 \label{ssinst}
\end{figure}

To better understand the instability we observed, we summarise the theory of SS instability below. When a suspended cavity is locked, the high power resonating light acts as an optical spring and couples the two mirrors. The cavity exhibits two different angular modes in the presence of any mirror misalignment, illustrated in Fig.~\ref{ssinst}. In one case, the radiation pressure stiffens the angular motion and hence to increase the angular frequency of the mirrors. In the other case, the radiation pressure softens the mirror angular motion and reduces the reduce the mirror angular frequency.

These two distinct modes of motion are called the cavity hard mode and soft mode, respectively. Without consideration of the control system, the hard mode is always stable since the radiation pressure induced torque is working in-phase with the natural pendulum mechanical restoring torque. However the soft mode can become unstable, as the radiation pressure on the suspended mirror works as an optical anti-spring on the mirror. If the radiation pressure torque is greater than the intrinsic restoring torque due to the local angular control system, the cavity soft mode frequency will be shifted to zero, and thus the cavity will be unstable.

Assuming that a simple FP cavity is slightly misaligned from its optimal position and the radiation pressure force of the off-centerd beam can be regarded as a point force. The differential equation that describes the dynamics of the suspended mirrors in this cavity is~\cite{angularsigg2003}
\begin{equation}\label{differential}
\bm{\ddot \theta} + \gamma \bm {\dot \theta} + \omega_{0}^2 \bm \theta+ \bm \Omega^2 \bm \theta = 0,
%\frac{d^{2}}{dt^{2}}\bm{\theta} = -\omega^2\theta_{i} - \bm{\Omega}_{i}^2\theta_{i},
\end{equation}
with,
\begin{equation}
\begin{split}
&\bm\theta = 
\left[ \begin{array}{c}
\theta_{1} \\
\theta_{2}
\end{array} 
\right ],\\
&\bm \Omega^2 = -\frac{2PL}{Ic}
\left [
        \begin{array}{c} 
	\frac{g_{2}}{1 - g_{1}g_{2}} \\ \frac{1}{1 - g_{1}g_{2}} 
	\end{array} 
	\begin{array}{c} 
	\frac{1}{1 - g_{1}g_{2}} \\ \frac{g_{1}}{1 - g_{1}g_{2}} 
	\end{array} 
\right ],
\end{split}
\end{equation}
where $\theta_{1,2}$ represent the angular motion of the two mirrors. The two mirrors have the same damping $\gamma$ and the angular resonant frequency $\omega_{0}$ since they are suspended and controlled in the same way. The $\bm{\Omega^2}$ term represents the radiation pressure effect on the mirror where $P$ represents intra-cavity power, $L$, the cavity length, $I$, the moment of inertia of the cavity mirrors, and $c$, the speed of light. The $g$-factor is defined by $g_{1,2} = 1 - L/R_{1,2}$ with $R_{1,2}$ representing the radius of curvature of two mirrors.

%of The first part in the right hand side of Eq.~\ref{differential} represents the natural pendulum mechanical restoring torque, while the second part is the radiation pressure induced torque of the suspended mirrors which can be written as:

%\begin{equation}\label{torque}
%\frac{d^{2}}{dt^{2}}\theta_{1} = -\omega_{1}^2\theta_{1} + \frac{2PL}{Ic}(\frac{g_{2}}{1 - g_{1}g_{2}}\theta_{1} + \frac{1}{1 - g_{1}g_{2}}\theta_{2})
%\end{equation}
%\begin{equation}\label{torque}
%\frac{d^{2}}{dt^{2}}\theta_{2} = -\omega_{2}^2\theta_{2} + \frac{2PL}{Ic}(\frac{1}{1 - g_{1}g_{2}} 
%\theta_{1} + \frac{g_{1}}{1 - g_{1}g_{2}}\theta_{2})
%\end{equation}
%where $P$ is intra-cavity power, $L$ is cavity length, $I$ is moment of inertia of cavity mirror, $c$ is speed of light. The $g$-factor for each mirror is defined as $g_{i} = 1 - L/R_{i}$ and $R_{i}$ is the radius of curvature of mirrors.

%From equation.1, we can write the requirement for stability as:
%\begin{equation}
%\Omega_{i}^2 > -\omega_{i}^2
%\end{equation}
The eigenvalues of $\bm{\Omega}$ can be written as
\begin{equation}\label{omega}
          \begin{aligned}
          \Omega_{\rm{h}} = [-\frac{PL}{Ic}\frac{g_{1} + g_{2} - \sqrt{4 + (g_{1} - g_{2})^2}}{1 - g_{1}g_{2}}]^{\frac{1}{2}},\\
          \Omega_{\rm{s}} = [-\frac{PL}{Ic}\frac{g_{1} + g_{2} + \sqrt{4 + (g_{1} - g_{2})^2}}{1 - g_{1}g_{2}}]^{\frac{1}{2}},
          \end{aligned}
\end{equation}
where the subscripts h and s of $\Omega$ are no longer representing the two individual mirrors but the angular modes of the cavity which we call hard and soft mode illustrated in Fig.~\ref{ssinst}.

The torsional spring constant of a suspended mirror, regarded as a torsional pendulum, is $K_{\rm{p}} = I \omega_{0}^2$, and the optical torsional spring constant due to radiation pressure in the suspended cavity is $K_{\rm{rp,i}} = I \Omega_{\rm{i}}^2$,~(i = h, s). %The radiation pressure torque works with or against the restoring torque provided by the natural torsional pendulum depending on the cavity angular mode. 
The total torsional spring constant $K_{\rm{total}}$ in the system is now modified to
\begin{equation}
K_{\rm{total,i}} = K_{\rm{p}}+K_{\rm{rp,i}},~(\rm{i} = h, s).
\end{equation}
$K_{\rm{total}}$ determines the stability of the pendulum angular modes. If $K_{\rm{total}} > 0$, there will always be a restoring torque to bring the torsional spring to its equilibrium position. However, if $K_{\rm{total}} < 0$, the suppressed restoring torque will generate an anti-torsional-spring, causing the pendulum angular mode to become unstable.

We know that in order to sustain stable Gaussian beam resonance in a simple two-mirror cavity, the $g$-factors are confined by $0< g_{1}g_{2}<1$. If we apply this criterion to Eq.~(\ref{omega}), we can determine that $K_{\rm{rp,h}}$ is always positive and $K_{\rm{rp,s}}$ is always negative. The sign of $K_{\rm{rp}}$ indicates whether the radiation pressure torque is working with or against the original restoring torque. The cavity hard mode corresponds to $K_{\rm{rp,h}}$ is statically stable. The soft mode associated with $K_{\rm{rp,s}}$ has the potential of becoming unstable, as radiation pressure reduces the total stiffness of the torsional pendulums. SS instability occurs when $K_{\rm{rp,s}} \leqslant -K_{\rm{p}}$. This only depends on the intra-cavity power $P$ since other parameters are constants for a cavity whose geometry is fixed. The unstable threshold power $P_{\rm{thr}}$ can be expressed as
\begin{equation}\label{threshold}
P_{\rm{thr}} = \frac{I\omega_{0}^2c(1-g_{1}g_{2})}{L\left[g_{1}+g_{2}+\sqrt{(g_{1}-g_{2})^2+4}\right]}.
\end{equation}

%The two eigenvalues of $K_{rp}$ are the optical spring constant, together with the pendulum mode, they form the cavity soft and hard modes. Their resonant frequencies can be written as:
%\begin{equation}
   %      \begin{aligned}
      %   f_{hard} = \frac{1}{2\pi}\sqrt{\frac{K_{pend} + K_{hard}}{I}}\\
         %f_{soft} = \frac{1}{2\pi}\sqrt{\frac{K_{pend} + K_{soft}}{I}}
         %\end{aligned}
%\end{equation}

%For the cavity hard mode, $K_{hard}$ is positive and the two torques act in the same direction, $f_{hard}$ increases and the mode is stable. When the radiation torque is larger than the pendulum restoring torque and acts in the opposite direction, $K_{soft}$ becomes negative. $f_{soft}$ is imaginary and the cavity soft mode is unstable.

If we apply this SS instability criterion to our Gingin cavity with the parameters listed in the Table.~\ref{Gingincavityparameters}, the threshold power for yaw mode is $P_{\rm{thr,yaw}}$ = 305.5 kW. 
%The threshold power for yaw and pitch are $P_{\rm{threshold,yaw}}$ = 305.5 $kW$ and $P_{\rm{threshold,pitch}}$ = 3.32 $MW$, respectively.
\begin{table}[h]
\caption{Gingin cavity parameters}
\centering
     \begin{tabular}{c | c } 
     \hline
     \hline
      Parameters     &     Value \\
     \hline
      Mirror mass $m$& 0.8 kg \\
      Yaw mode frequency $f_{\rm{yaw}}$& 1.7 Hz \\
      %Pitch mode frequency $f_{\rm{pitch}}$& 5.6 $Hz$\\
      Moment of inertia $I$ (in yaw) & 6.67 $ \times 10^{-4}~\rm{kg \cdot m^{2}}$\\
      Cavity length $L$  & 74 m   \\
      ITM curvature $R_{1}$ & 37.3 m \\
      $g_{1}$ &  -0.9839\\
      ETM curvature $R_{2}$ & 37.4 m \\
      $g_{2}$ & -0.9786\\
     \hline
     \end{tabular}
     \label{Gingincavityparameters}
\end{table}

\section{Observation of angular instability in Gingin cavity}
\begin{figure}[h!]
   \centering
  % Requires \usepackage{graphicx}
   \includegraphics[width=9cm]{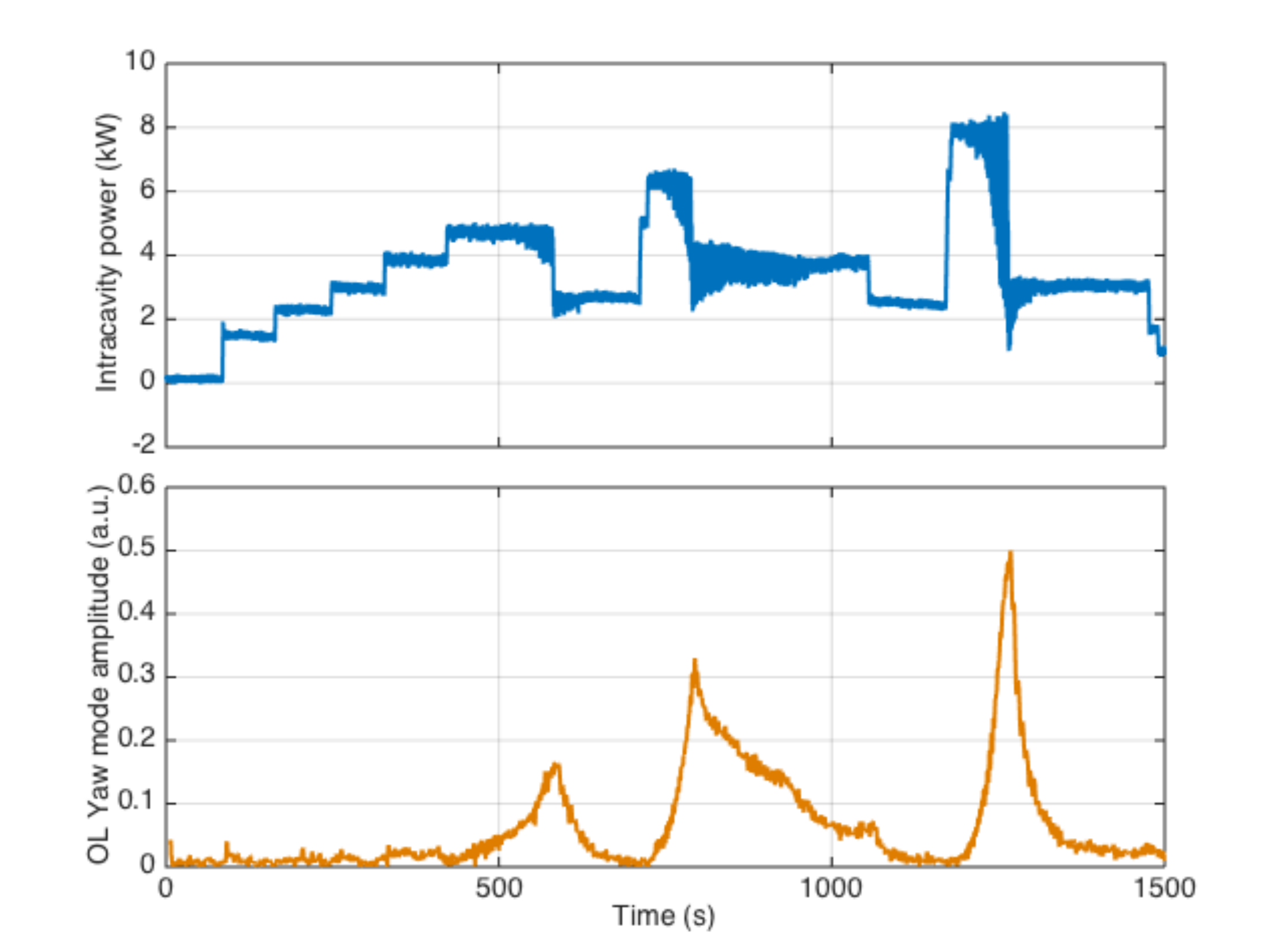}\\
\caption{Observation of angular instability in Gingin cavity. The upper panel shows a 25 minutes measurement of intra-cavity power. The lower panel shows the amplitude of yaw 1.7 Hz mode measured from the optical lever. When the power reaches a threshold, the yaw mode begins to ring up. Subsequent lowering of the intra-cavity power causes the mode starts to ring down. Three different yaw mode ring-ups occur at different intra-cavity power levels. The ring up is faster when the intra-cavity power is higher.}
\label{ringup}
\end{figure}

Theoretical calculations have shown that the SS angular instability threshold power for our mirror yaw mode is about 300 kW. However, we observed an angular instability at only 1.3\% of this threshold. By monitoring the 1.7 Hz yaw mode amplitude from the optical lever and the intra-cavity power from the cavity transmission photo detector, it was observed that when the intra-cavity power exceeded a threshold power about 4 kW, the yaw mode rang up exponentially, as seen in Fig.~\ref{ringup}. 

The time constant of the yaw mode amplitude ring-up depends on the intra-cavity power. The data of ring-up curves shown in Fig.~\ref{ringup} was filtered with a 1.7 Hz band pass filter to isolate the yaw mode. The filtered ring-up curves were then fitted with an exponential function to obtain the ring-up constants ,$\tau$, for different intra-cavity powers. The ring-up time and the associated intra-cavity power were then fitted to determine threshold power using
\begin{equation}\label{ringupeq}
  \frac{1}{\tau}=a\sqrt{P}+b,
\end{equation}
where $\tau$ is the ring-up time constant and $P$ is the average intra-cavity power during the ring-up time. Parameters $a$ and $b$ are related to the local control of the yaw mode, which can be fitted with respect to the ring-up time and intra-cavity power data. The threshold power is the power when the ring-up time is infinite, which is $P_{\rm{thr}}=(-b/a)^{2}$. In our case $a = 0.064 \pm 0.096$, $b = - 0.126 \pm 0.208$ with 95\% confidence and the fitting $\rm{R^2}$ is 0.994, indicating that $P_{\rm{thr}}$ is about 4 kW.

This low threshold power angular instability sets a limit on the intra-cavity power, which in turn limits our ability to work on experiments studying parametric instability that require higher optical power. Moreover, this is not the original SS angular instability, because the intra-cavity power is significantly lower than the SS angular instability threshold power as calculated in the previous section. The instability we observed here does not arise from the soft mode frequency reduced to zero. Instead, we observed an instability that arises from the interaction between the SS effects on the soft and hard modes; and the mirror control loops.%The threshold power could be increased by improving the local angular control system to enhance the stiffness of the cavity soft mode.
%This leads us to construct a model of the yaw mode control system that includes both the local control loops and optical torsional spring effect.

\section{Angular control system} 

% our control system (our system setup + our system transfer function)+ observation of the instability + modified transfer function.

The mirrors of our cavity are suspended from the control mass stage via four high quality factor niobium wire-flexure modules~\cite{fang2018modular}. This control mass stage is suspended from a three-stage pendulum, which is in turn suspended from a 3D low frequency pre-isolator. An optical lever is used for directly sensing of the mirrors pitch and yaw motion. While several shadow sensors located at the control mass provide additional sensing. These sensing signals are processed by a digital signal processor with built-in filter banks and fed back to the coil-magnet actuators to alter the control mass in different degrees of freedom. In this way the static position of the optic is maintained and the mirror suspension pitch and yaw modes are damped. In this paper we consider the yaw mode in detail and assume that there is no coupling between different degrees of freedom. 

\begin{figure}[h!]
          \centering
          \includegraphics[width=9cm]{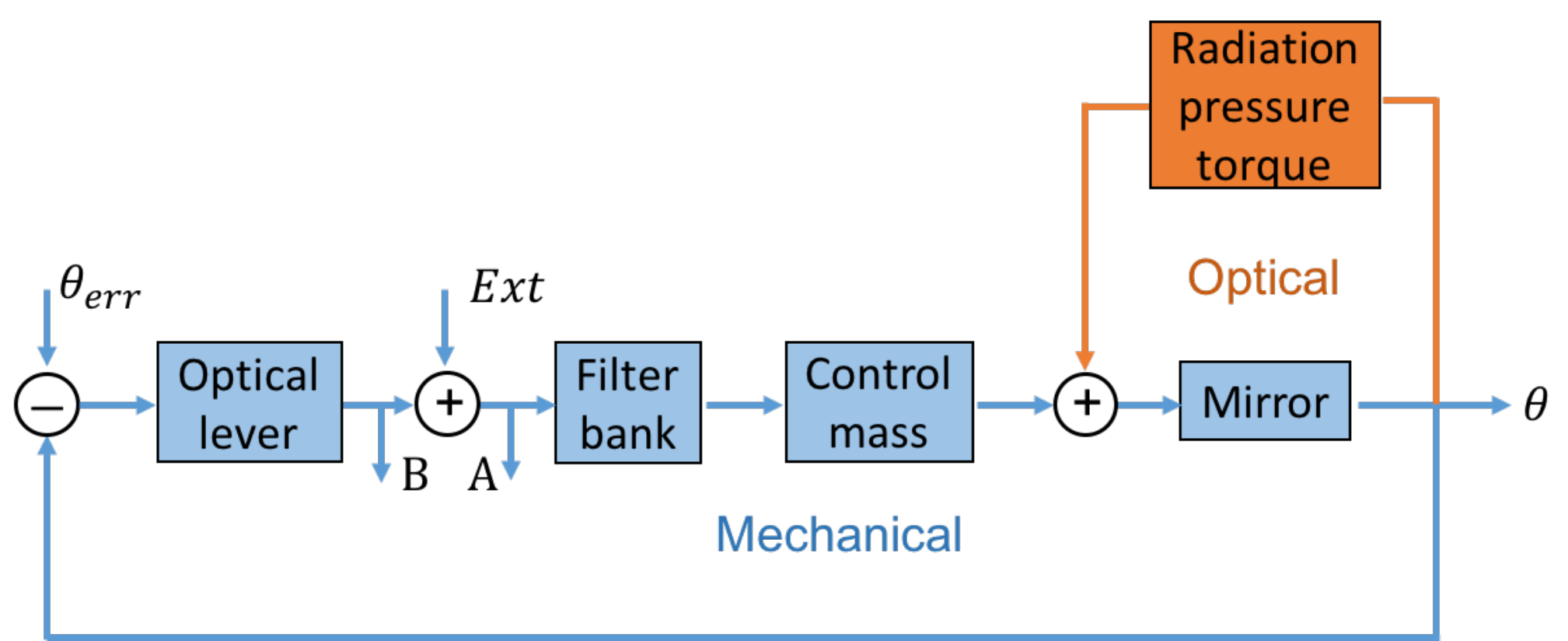}\\
          \caption{The block diagram of mirror yaw mode control system. The blue parts are the subsystems that involved in the mechanical control. The orange part is the optical control which is only valid when the cavity is locked. An external excitation signal is injected and added to the optical lever signal for measuring the control open loop transfer function (OLTF).}
\label{block}         
\end{figure}

The same angular control is applied locally on both suspended mirrors. The feedback control schemes are necessary to damp mirror motion in order to easily achieve cavity locking. When the cavity is not locked, the two mirrors' local control systems are independent and stable. If the cavity is locked, the two mirrors are coupled by radiation pressure. There will be a resultant radiation pressure torque acting as a parallel control loop and adding to the local control, illustrated as the orange curve in Fig.~\ref{block}. This would modify the mirror angular control loop transfer function.

The yaw mode control block diagram in our system is illustrated in Fig.~\ref{block}. The mechanical control part (blue) is always applied and the radiation pressure effect part (orange) is only valid when the cavity is locked.

We first measured the open loop response of the mirror yaw local mechanical control system without locking the cavity. We injected an external excitation signal ($Ext$) to add with the original optical lever signal. The summation signal is sent to the control system for angular control. We simultaneously recorded signals from two test points A and B which are located just upstream and downstream of the excitation signal. With the cavity unlocked, we can measure the OLTF of the control system in the absence of radiation pressure effect, which can be written as
\begin{equation}\label{eq7}
          G_{\rm{local}} = \frac{B}{A},  
\end{equation}
where the $G_{\rm{local}}$ is the product of transfer functions of several subsystems such as the photo-detector response, actuator response, filter banks and the effects of the upper three-stage pendulums. 
\begin{figure}[h!]
\hspace*{0in}
\centering
\includegraphics[width=8cm]{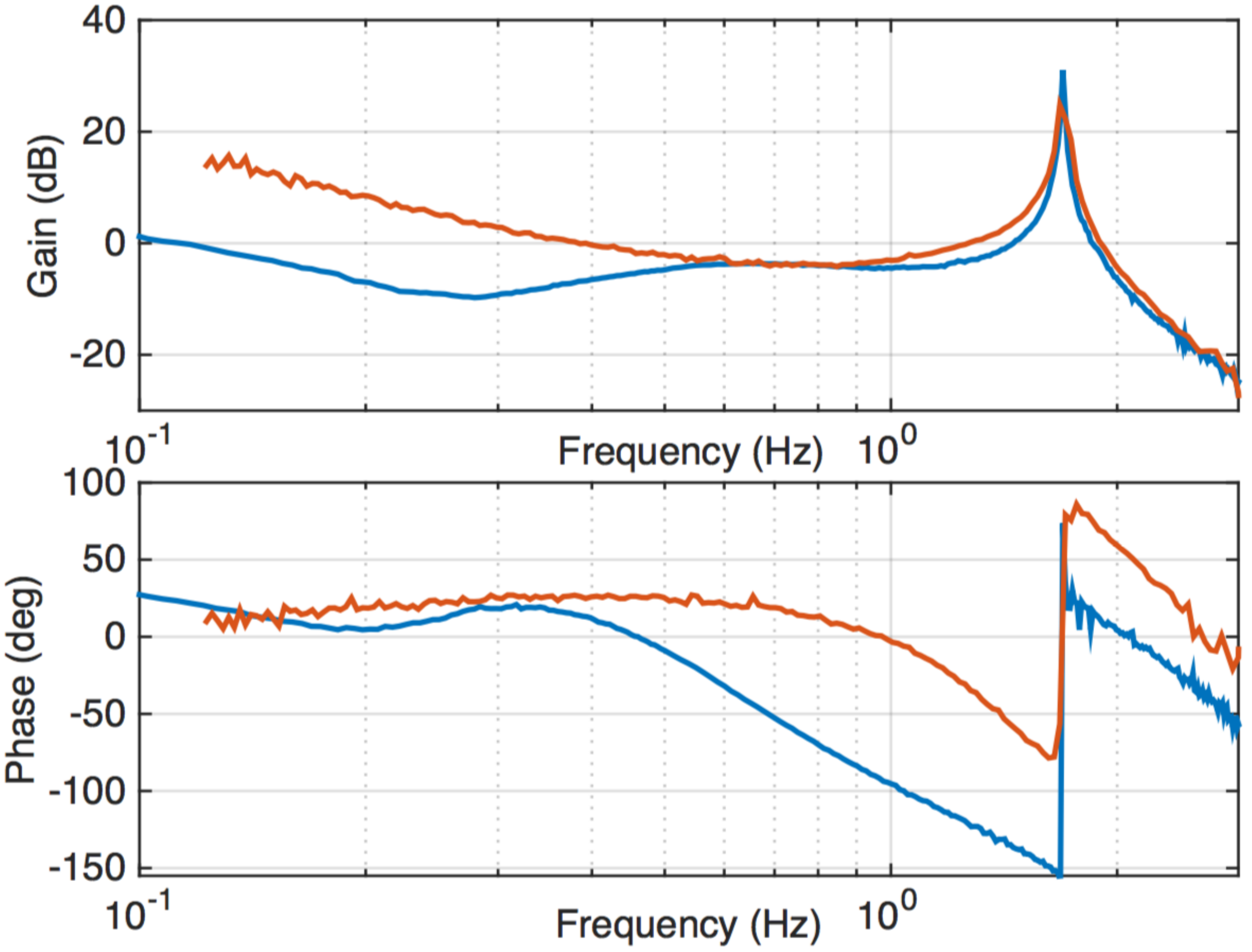}
\caption[Gingin mirror yaw motion control loop open loop transfer function]{Gingin mirror yaw motion control loop OLTF. The blue curve is the yaw OLTF measured when we observed the yaw angular instability with about 4 kW threshold power. The orange curve is the yaw OLTF after we changed the filter settings and improved the control.}
\label{tfcompare}
\end{figure}

The original OLTF is the blue curve in Fig.~\ref{tfcompare} %Peaks at low frequency are the resonances of the multi-stage pendulums. %We don't have high control gain at 21 mHz and 38 mHz, oscillations at these frequencies take very long time to damp. 
with the peak at 1.7 Hz being the mirror yaw mode. %we need to have huge gain at this frequency to suppress the noise at this resonance frequency.
Two unity gain frequencies (UGFs) are located around the resonance peak at 1.45 Hz and 1.85 Hz with phase of -136 degrees and 15 degrees respectively. By changing the control gain settings and bandwidth and frequency of filters in the digital filter bank, we obtained the improved OLTF with enhanced phase margin, as illustrated in the orange curve of Fig.~\ref{tfcompare}. As it can be seen, the new UGFs are now 1.27 Hz and 1.90 Hz and the new system phase margin is about 40 degrees. 

\begin{figure}[h!]
\hspace*{0in}
\centering
\includegraphics[width=8cm]{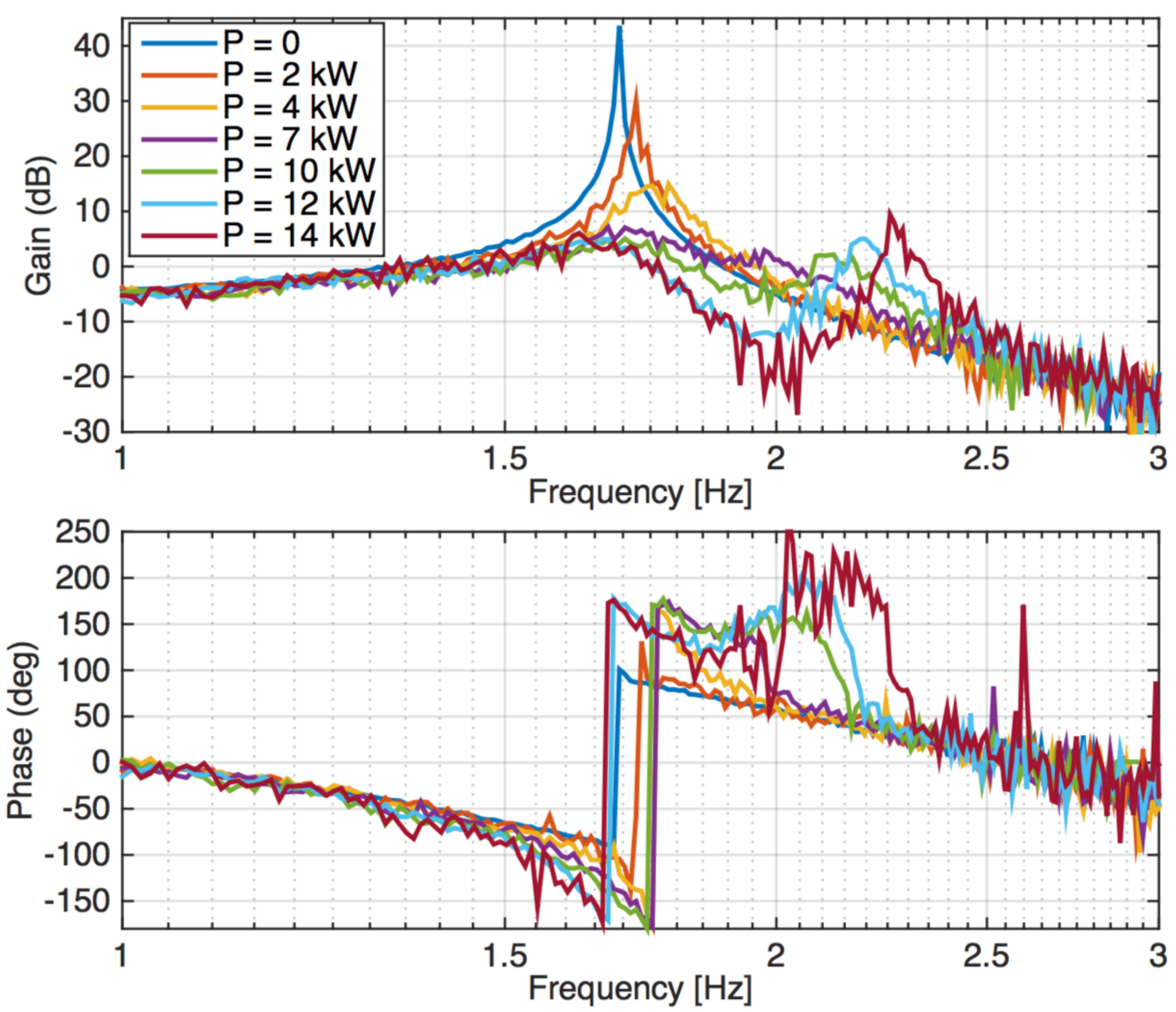}
\caption{Radiation pressure effect modified yaw control OLTF. The external drive signal is kept at a low level to maintain stable cavity locking at the expense of reducing coherence.}
\label{RPmodification}
\end{figure}

When the cavity is locked, the optical part in the block diagram becomes valid. We measured the radiation pressure modified OLTF and found that after the improvement illustrated in Fig.~\ref{tfcompare}, our cavity stability is now no longer limited by 4 kW threshold power. We increased the intra-cavity power up to 14 kW, the cavity was well locked and stable. This means that the angular instability we observed previously is caused by the poor phase margin of the original control loop and radiation pressure effect reduces the system phase margin. We measured the OLTF with different intra-cavity power levels and the results are shown in Fig.~\ref{RPmodification}.

The interaction between local control loop and radiation pressure effect can be clearly seen. When the cavity is locked, the two mirrors are coupled and the yaw control loop is no longer affecting individual mirror locally, but is instead acting on cavity angular modes as described in the SS model. As the power increases, the yaw mode splits into two which are interpreted to be the cavity soft and hard modes. The soft mode shifts to lower frequency slowly and will theoretically be shifted to zero at 305 kW, whilst the hard mode shifts to higher frequency more rapidly as intra-cavity power increases.  If the power is high enough, the gain of soft mode peak will be below 0 dB. However the gain of hard mode peak will be above unity gain, assuming a reasonable phase margin the system will be stable. 

If we look carefully at the phase curve in Fig.~\ref{RPmodification}, we can find out that the OLTF phase curves of different powers overlap with each other after the higher frequency UGF of the hard mode, meaning the phase margin will still follow the original phase curve. This indicates that the phase margin will be zero again if the UGF reaches 2.7 Hz. The added interaction due to radiation pressure modifies the control loop such that the phase margin is reduced, making our system susceptible to instability. If we want our system to be less vulnerable to high power angular instability, we need to have a larger phase margin at higher frequencies, or modify the control filters as the power is increased.

\section{Simulation}
\begin{figure*}[t!]
\centering
\subfigure[Full Simulink model for radiation pressure effect simulation.] {\includegraphics[width=14cm]{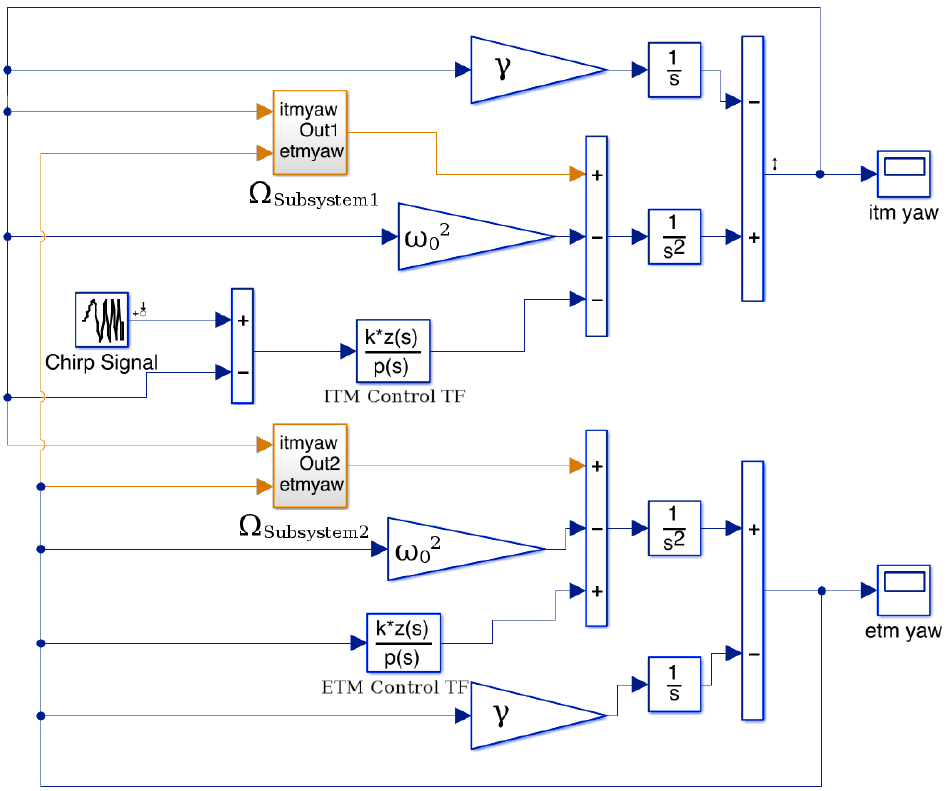}}
\subfigure[$\bm\Omega_{\rm{Subsystem1}}$] {\includegraphics[width=8cm]{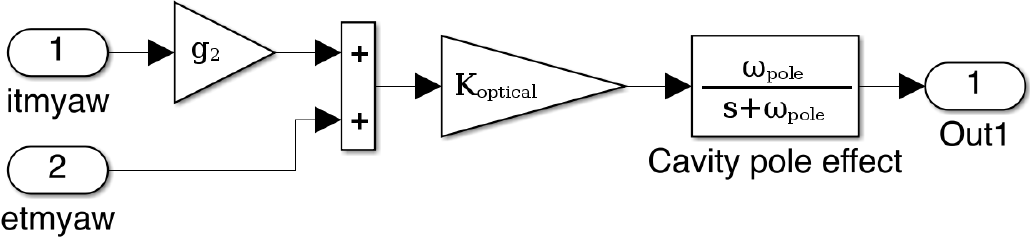}}
\subfigure[$\bm\Omega_{\rm{Subsystem2}}$] {\includegraphics[width=8cm]{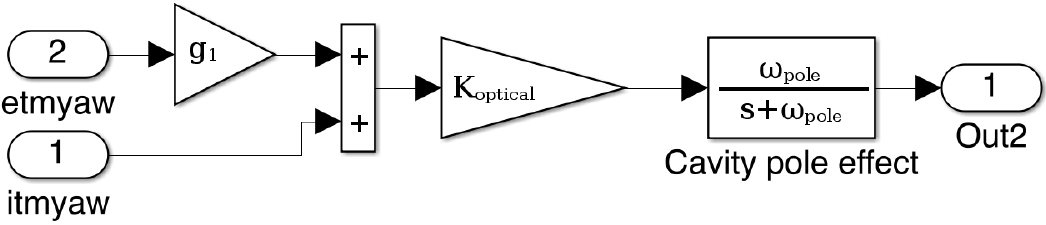}}
\caption{Simulink model for radiation pressure effect simulation. Blue lines indicate local controls for the ITM and ETM mirrors, and orange lines indicate optical radiation pressure coupling components. $K_{\rm{optical}} = \frac{2PL}{cI(1-g_{1}g_{2})}$ in the subsystems determines the optical torsional spring strength. The cavity pole effect is also included in this subsystem model.}
\label{simulink}
\end{figure*}

To analysis the radiation pressure effect on the control system and predict the unstable power threshold, we simulated the system response to radiation pressure using SimuLink,  enabling us to analyse the system in both time and frequency domain. The completed SimuLink model is shown in Fig.~\ref{simulink}. We first fit our mechanical control open loop measurement data to a continuous transfer function model. The transfer function model blocks in Fig.~\ref{simulink} represents the local controls. Without laser power, the whole model can be simplified into two independent parts for two mirrors (named ITM and ETM) individually. In our model, both mirrors use the same transfer function block for local control since the measured open loop response for both mirrors are actually the same. In this model the radiation pressure effect parts are represented using two subsystems. The two subsystems couple individual mirror angular motion together and form cavity soft and hard modes which are as described in the SS instability model. 

A chirp signal is added with ITM yaw which is the same as the $Ext$ signal shown in Fig.~\ref{block}. Control torque and radiation pressure torque are added with mirror torsional pendulum natural restoring torque to create the new mirror response. Damping is defined according to the pendulum Q-factor (we use Q = 100 in the simulation). Two Simulink Linear Analysis points are allocated to the chirp signal and ITM yaw to help us obtain the frequency response of the system. Simulink Linear Analysis Tool gives the closed loop response bode diagram of the system at different values of optical power input. The closed loop transfer function is converted to an OLTF to compare with the measurements.
\begin{figure}[h!]
\hspace*{0in}
\centering
\includegraphics[width=8cm]{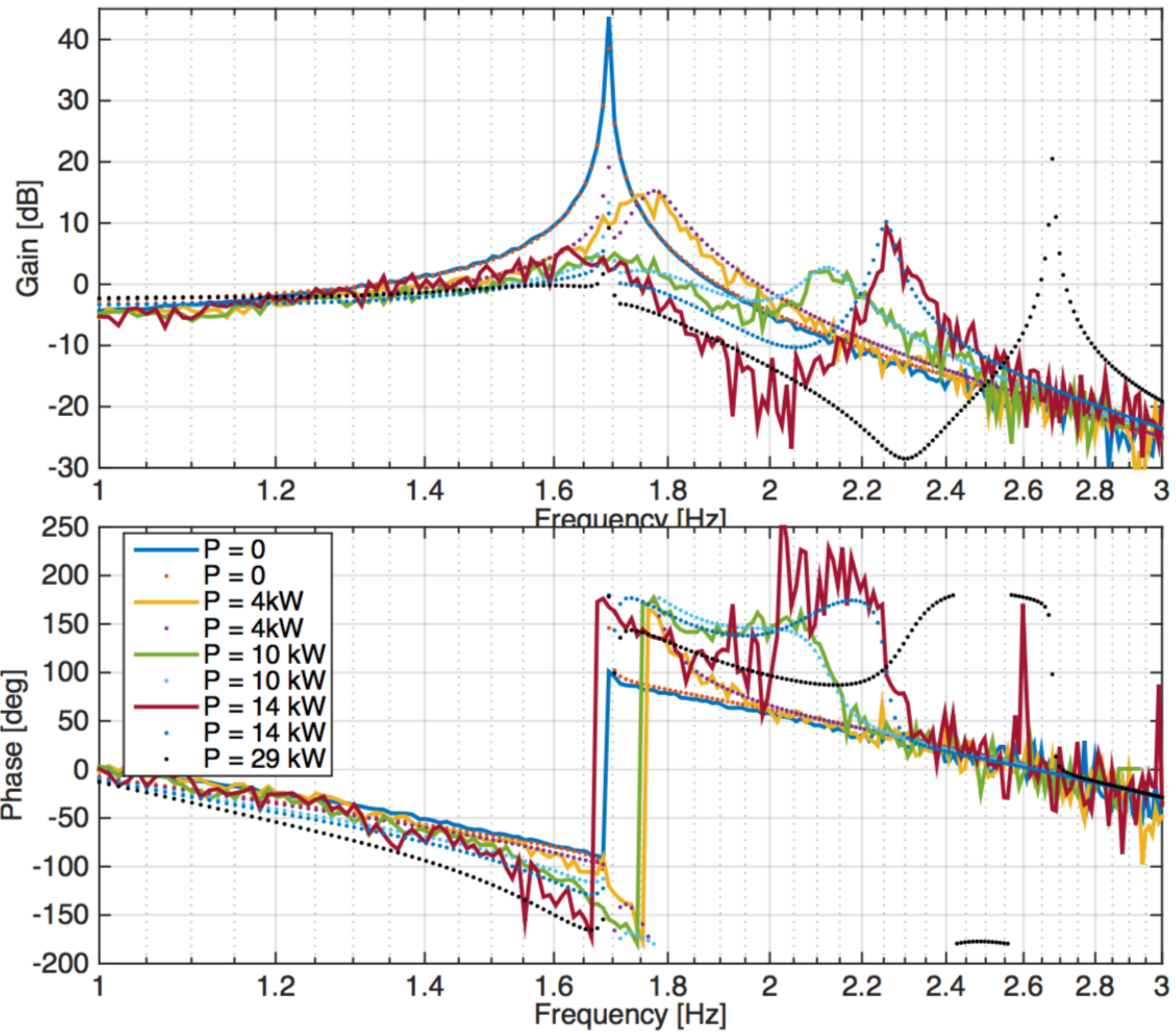}
\caption{Simulink fitting results of Gingin angular instability. The solid curves are measurement data and the dotted curves are simulink fitting. The measurement curves are selected from Fig.~\ref{RPmodification} to show the trend of radiation pressure modification on the loop transfer function. }
\label{fitting}
\end{figure}  

Simulation results fit with the measurement data quite well. The simulation also predicts the threshold power will be 29 kW, which is shown in the black dotted curve in Fig.~\ref{fitting}. The phase at UGF (around 2.7 Hz) in the curve is about zero, this is consistent with our prediction that the system will be unstable when the phase at UGF is reduced to zero due to interaction between local control and radiation pressure effect.

\section{Conclusion}

We have shown that the dynamics of a suspended cavity is modified by the radiation pressure forces which can lead to an angular instability at a much lower intra-cavity power level than the SS instability. This new type of angular instability was observed in a 74 m suspended optical cavity. The interaction between the local control loop and radiation pressure alters the local control loop transfer functions in such a way that the phase margin is reduced and causes an instability. 

Simulation results based on Simulink fit with the measurement data well and the threshold power can be predicted. It is also possible to empirically predict how prone a cavity is to this instability from direct OLTF measurement, by observing the zero phase frequency that the hard mode will eventually reach. The instability threshold power can be enhanced by improving the local control phase margin. In our system, optimising the local control loop phase increases the threshold power from 4 kW to about 29 kW. 

aLIGO-like suspended cavities with negative $g$-factor are intrinsically more susceptible to the reported instability, but careful design and tuning of the local angular control loops will help achieve higher threshold power. For aLIGO and future detectors, if we don't take this effect into consideration in the first place and modify the controls according to the intra-cavity power level, instability would occur much earlier. 

When the hard mode frequency shifts to close to the detection band, the angular control loop bandwidth will need to be increased to the detection band in order to maintain stability. Unfortunately, the broadband angular control loop would couple extra noise from the angular sensing. It is necessary to keep the hard mode frequency much lower than the detection band frequencies by increasing the test mass size or weight. 

For current aLIGO with $\sim$ 200 kW circulating power, the hard mode frequency will be $\sim$ 2 Hz, it should not be a problem to optimise the control loop to maintain the stability. When operating at the design power $\sim$ 800 kW, the hard mode frequency will shift to $\sim$ 4 Hz, it will be more difficult to keep it stable without noise injection at $\sim$ 10 Hz.

This work would provide useful information for angular control loop design for future gravitational wave detectors.

\section*{Acknowledgments}
We wish to thank the LIGO Scientific Collaboration Advanced Interferometer Configurations and Optics Working Group for useful advice. We also want to thank Joris van Heijningen for helpful comments on improvement of the manuscript. This work is supported by the Australian Research Council Centre of Excellence for Gravitational Wave Discovery project CE170100004.

% Create the reference section using BibTeX:

\section*{References}
\bibliographystyle{unsrt}
%\bibliography{refs}

% Run this once to generate your BBL file. Then copy the contents of your BBL file into your main latex file, commenting out "\bibliography"

\end{document}